\documentclass[aps,superscriptaddress,reprint]{revtex4-1}
\usepackage[english]{babel}
\usepackage{amsmath,amsthm,amssymb}
\usepackage{amsfonts}
\usepackage{xspace}
\usepackage{bm}
\usepackage{graphicx}
\usepackage[separate-uncertainty = true, exponent-product = \times]{siunitx}
\usepackage{hyperref}
\usepackage{dcolumn}
\usepackage{color}

\newcommand{\red}[1]{\textcolor{black}{#1}}

\begin{document}

\title{Sign of inverse spin Hall voltages generated by ferromagnetic
resonance and temperature gradients in yttrium iron garnet{\textbar}platinum bilayers}
\author{Michael Schreier}
\email{michael.schreier@wmi.badw.de}
\affiliation{Walther-Mei\ss ner-Institut, Bayerische Akademie der Wissenschaften,
Garching, Germany}
\affiliation{Physik-Department, Technische Universit\"at M\"unchen, Garching, Germany}

\author{Gerrit~E.~W. Bauer}
\affiliation{Institute for Materials Research, Tohoku University, Sendai, Japan}
\affiliation{WPI Advanced Institute for Materials Research, Tohoku University,
Sendai, Japan}
\affiliation{Kavli Institute of NanoScience, Delft University of Technology, Delft, The
Netherlands}

\author{Vitaliy I. Vasyuchka}
\affiliation{Fachbereich Physik and Landesforschungszentrum OPTIMAS, Technische Universit\"at Kaiserslautern, Kaiserslautern,
Germany}

\author{Joost Flipse}
\affiliation{Physics of Nanodevices, Zernike Institute for Advanced Materials, University
of Groningen, Groningen, The Netherlands}

\author{Ken-ichi Uchida}
\affiliation{Institute for Materials Research, Tohoku University, Sendai, Japan}
\affiliation{PRESTO, Japan Science and Technology Agency, Saitama, Japan}

\author{Johannes Lotze}
\affiliation{Walther-Mei\ss ner-Institut, Bayerische Akademie der Wissenschaften,
Garching, Germany}
\affiliation{Physik-Department, Technische Universit\"at M\"unchen, Garching, Germany}

\author{Viktor Lauer}
\affiliation{Fachbereich Physik and Landesforschungszentrum OPTIMAS, Technische Universit\"at Kaiserslautern, Kaiserslautern,
Germany}

\author{Andrii V. Chumak}
\affiliation{Fachbereich Physik and Landesforschungszentrum OPTIMAS, Technische Universit\"at Kaiserslautern, Kaiserslautern,
Germany}

\author{Alexander A. Serga}
\affiliation{Fachbereich Physik and Landesforschungszentrum OPTIMAS, Technische Universit\"at Kaiserslautern, Kaiserslautern,
Germany}

\author{Shunsuke Daimon}
\affiliation{Institute for Materials Research, Tohoku University, Sendai, Japan}

\author{Takashi Kikkawa}
\affiliation{Institute for Materials Research, Tohoku University, Sendai, Japan}

\author{Eiji Saitoh}
\affiliation{Institute for Materials Research, Tohoku University, Sendai, Japan}
\affiliation{WPI Advanced Institute for Materials Research, Tohoku University,
Sendai, Japan}
\affiliation{Advanced Science Research Center, Japan Atomic Energy Agency, Tokai, Japan}
\affiliation{CREST, Japan Science and Technology Agency, Tokyo, Japan}

\author{Bart~J. van~Wees}
\affiliation{Physics of Nanodevices, Zernike Institute for Advanced Materials, University
of Groningen, Groningen, The Netherlands}

\author{Burkard Hillebrands}
\affiliation{Fachbereich Physik and Landesforschungszentrum OPTIMAS, Technische Universit\"at Kaiserslautern, Kaiserslautern,
Germany}

\author{Rudolf Gross}
\affiliation{Walther-Mei\ss ner-Institut, Bayerische Akademie der Wissenschaften,
Garching, Germany}
\affiliation{Nanosystems Initiative Munich, Munich, Germany}
\affiliation{Physik-Department, Technische Universit\"at M\"unchen, Garching, Germany}

\author{Sebastian~T.~B. Goennenwein}
\affiliation{Walther-Mei\ss ner-Institut, Bayerische Akademie der Wissenschaften,
Garching, Germany}
\affiliation{Nanosystems Initiative Munich, Munich, Germany}

\begin{abstract}
We carried out a concerted effort to determine the absolute sign of the inverse spin Hall
effect voltage generated by spin currents injected into a normal metal. We focus on yttrium iron garnet (YIG)$|$platinum bilayers at room temperature, generating spin currents by microwaves and temperature gradients. We find consistent results for different samples and measurement setups that agree with theory. We suggest a right-hand-rule to
define a positive spin Hall angle corresponding to the voltage expected for the simple case of scattering of free electrons from repulsive Coulomb charges.%
\end{abstract}

\maketitle

The bon mot that the sign is the most difficult concept in physics since
there are no approximate methods to determine it has been ascribed to Wolfgang Pauli.
Indeed, the struggle to obtain correct signs permeates all of physics. While
the negative sign of the electron charge is just a convention, that of
derived properties, such as the (conventional) Hall voltage, has real physical meaning. Often it is much easier and
sufficient to determine sign differences between related quantities. However, a
complete understanding requires not only the relative but also the absolute
sign. Here we address the sign of the (inverse) spin Hall effect
[(I)SHE]~\cite{Dyakonov1971,Hirsch1999,Zhang2000,Murakami2006,Sinova2004,Kato2004,Saitoh2006,Valenzuela2006, Kimura2007, Wunderlich2010}
and related phenomena. The characteristic parameter is the spin Hall angle,
defined as the ratio $\theta _\mathrm{SH}\propto J_\mathrm{s}/J_\mathrm{c}$ of the transverse spin current $J_\mathrm{s}$ caused by an
applied charge current $J_\mathrm{c}$ (a more precise
definition is given below). The sign of $\theta_\mathrm{SH}$ may differ for
different materials. Since the spin Hall angle for Pt is generally taken to
be positive, $\theta_\mathrm{SH}$ of Mo~\cite{Mosendz2010}, Ta~\cite{Liu2012}, and W~\cite{Pai2012} must be negative.

The sign of $\theta_\mathrm{SH}$ governs the direction of the spin transfer
torque on a magnetic contact relative to that caused by the Oersted
magnetic field induced by the same current $J_\mathrm{c}$~\cite{Liu2012,Pai2012}.
It also determines the sign of the induced transverse voltage in experiments in which the ISHE is used to detect spin currents~\cite{Saitoh2006}. This technique is now widely used to study spin current
injection by a magnetic contact, through ``spin pumping'' induced by
ferromagnetic resonance (FMR)~\cite{Tserkovnyak2002a, Mosendz2010, Ando2010,
Sandweg2011, Czeschka2011, Castel2012} or by temperature differences~\cite%
{Uchida2008, Uchida2010, Jaworski2010, Qu2013, Schreier2013a}
(``spin Seebeck effect'', SSE). 

However, the
pitfalls that can affect the determination of the sign of the $\theta_\mathrm{SH}$, such as the sign of the spin currents~\cite{Jonker2004} and magnetic field direction are often glossed over in experimental and theoretical
papers. Moreover, a mechanism for a sign
reversal of the longitudinal spin Seebeck effect has recently been proposed~\cite{Adachi2013}. A careful analysis of experimental results with
respect to the signs of of FMR and thermal spin pumping voltages
generated by the inverse spin Hall effect is therefore overdue.

In this letter we present the results of a concerted action to resolve the sign issue by
comparing experiments on microwave-induced spin pumping and spin Seebeck effect for a bilayer of the magnetic insulator
yttrium iron garnet (YIG) and platinum (Pt) at room temperature. Samples grown by different techniques have been used in four experimental setups at the Institute for Materials Research, Tohoku University (IMR), Technische Universit\"{a}t Kaiserslautern (UniKL) Zernike Institute for Advanced Materials, University of Groningen (RUG) and the Walther-Mei{\ss }ner-Institut in Garching (WMI). Considering the different
sample properties the variations in the magnitude of the
observed voltages is not surprising. However, all groups find identical signs for the
ISHE Hall voltages that agree with the standard theory for spin
pumping by FMR~\cite{Tserkovnyak2002} and spin Seebeck effect~\cite%
{Xiao2010,Adachi2011,Hoffman2013}. A positive spin
Hall angle can be associated with scattering at negatively charged
Coulomb centers in the weakly relativistic electron gas.

Let us define the the electron charge as $-e<0.$ We recall that the thumb of the right hand points along the angular momentum $\mathbf{L}=\mathbf{r}\times \mathbf{p}$ of a
circulating particle with mass $m$, position $\mathbf{r}$ and momentum $%
\mathbf{p}=m\mathbf{v}$ when the tangential velocity $\mathbf{v}$ is along the fingers of its fist. The magnetic moment of a particle with charge $q$ is given by $\pmb{\mu}_L=q(2m)^{-1}\pmb{L}$~\cite{Jackson1975}. This magnetic moment direction is also generated
by two monopoles on the $\mathbf{L}$-axis, the negative (south pole) just
below and the positive (north) pole just above the origin. The magnetic
moment of the Earth points to the south, so the geographic north pole is
actually the magnetic south pole. Hence, the geomagnetic fields on the
surface of the Earth as measured by a compass needle point to the north
pole. The intrinsic angular momentum (spin) of non-relativistic electrons 
\cite{Schiff1949} is $\mathbf{s}=\frac\hbar2 \pmb{\sigma}$, where $%
\boldsymbol{\sigma }$ is the vector of Pauli spin matrices. The corresponding
magnetic moment $\pmb{\mu}_s=-e(2m)^{-1}g\pmb{s}=-\gamma 
\pmb{s}$, where $g$ is the g-factor and $\gamma$ is the gyromagnetic ratio.
In most solids $g$ and therefore $\gamma$ are positive.

\begin{figure}%[tbp]
\includegraphics[width=\columnwidth]{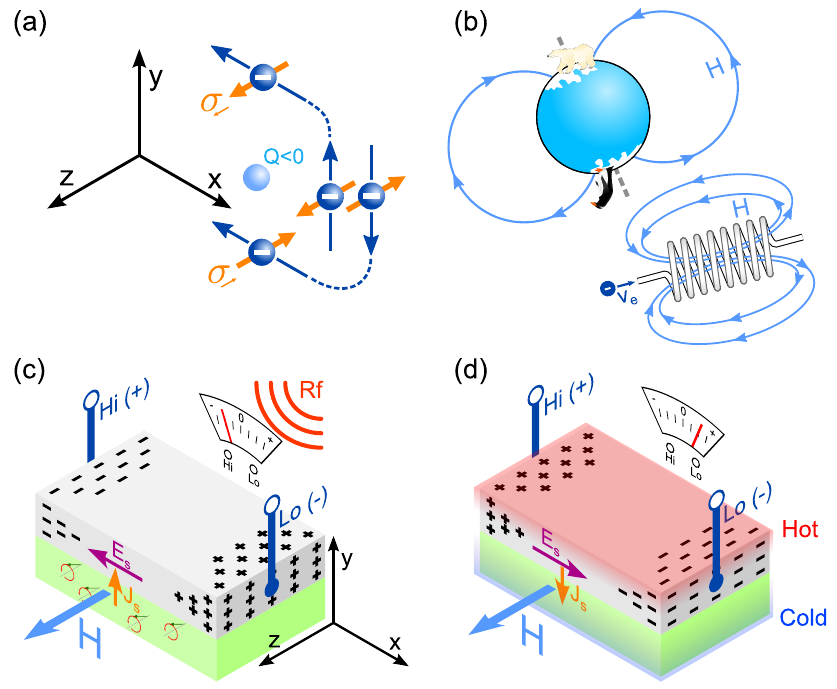}
\caption{(a) The transverse deflection of polarized electrons generated by a
fixed point charge $Q<0$ that we associate with a positive spin Hall angle.
(b) A magnetic field is positive when aligned with the Earth's magnetic
field or as the magnetic field generated by a negatively charged particle current flowing through a coil
when configured as sketched. [(c),(d)] Typical setups for spin pumping and
spin Seebeck experiments, respectively. [(c)] An rf-microwave field excites
magnetization precession that relaxes by emitting a spin current into the
adjacent Pt layer. [(d)] The spin current from the YIG to
the Pt is negative when the latter is hotter. In both cases, the ISHE
leads to a voltage between the contacts \textit{Hi} and \textit{Lo}.}%
\label{fig:defplusminus}
\end{figure}

The angular momentum and spin current tensor $\overleftrightarrow{\mathbf{J}}_\mathrm{s}$ consists of column vector elements $\mathbf{J}_\mathrm{s}^{\alpha }$ that represent the polarization of angular
momentum currents in the Cartesian $\alpha $-direction, while the row
vectors $\mathbf{J}_{\mathrm{s},\beta }$ represent the flow direction of angular
momentum along $\beta .$ The charge- ($\pmb{J}_\mathrm{c}$) and spin currents are both defined as particle flow (units $s^{-1}$). We can then \emph{define} the spin Hall angle $\theta _\mathrm{SH}$ as the proportionality factor in the phenomenological relations%
\begin{equation}
\mathbf{J}_{\mathrm{s},\beta }=\theta _\mathrm{SH}\pmb{\hat{\beta}}\times \mathbf{J}_\mathrm{c}
\label{eq:Js}
\end{equation}%
\begin{equation}
\mathbf{J}_\mathrm{c}=\theta _\mathrm{SH}\sum_{\alpha }\mathbf{J}_\mathrm{s}^{\alpha }\times 
\pmb{\hat{\alpha}}
\label{eq:ISHE}
\end{equation}
where the $\pmb{\hat{\alpha}},\pmb{\hat{\beta}}\in \left\{ \mathbf{\hat{x}},\mathbf{\hat{y}}%
,\mathbf{\hat{z}}\right\} $ are Cartesian unit vectors.

We now demonstrate the physical
significance of the sign of $\theta_\mathrm{SH}$. 
\red{We do not intend to model the material dependence or contribute to the discussion on whether observed effects are intrinsic or extrinsic.} %\red{We would like to stress that the model developed in the following is neither intended as quantitative theory nor aimed at accurately describing the microscopic origin of the spin Hall effect in real materials. Its sole purpose is to give a simple, comprehensible definition of the sign of the spin Hall angle based on the spin orbit interaction.} 
For an external point charge $Q$
at the origin in the weakly relativistic electron gas the bare Coulomb
potential at distance $r$ is%
\begin{equation}
\phi _{0}(r)=\frac{1}{4\pi \epsilon _{0}}\frac{Q}{r},  \label{eq:Epot}
\end{equation}%
where $\epsilon_{0}$ is the vacuum permittivity. In metals $\phi_{0}(r)$
is screened by the mobile charge carriers to become the Yukawa potential $%
\phi =\phi _{0}e^{-r/\lambda}$. The screening length $\lambda $ serves to
regularize the expectation values, but drops out of the final results. The
spin-orbit interaction of an electron in a potential $\phi $ is equivalent
to an effective magnetic field~\cite{Engel2007} 
\begin{equation}
\mathbf{B}_{\mathrm{so}}=\frac{-e}{2m^{2}c^{2}\gamma}\left( \boldsymbol{\nabla }\phi \times \mathbf{p}\right) ,  \label{eq:spin-orbit}
\end{equation}%
where $c$ is the velocity of light. The force on the electron then reads 
\begin{equation}
\mathbf{F}_{\mathrm{so}}=\boldsymbol{\nabla }(\boldsymbol{\mu }\cdot \mathbf{B}_{\mathrm{so}})=\frac{e \hbar}{4m^{2}c^{2}}\boldsymbol{\nabla }\left[ \boldsymbol{\sigma }\cdot \left( \boldsymbol{\nabla }\phi \times \mathbf{p}\right) \right] .  \label{eq:soforce}
\end{equation}%
Focussing on a free electron moving in the $y$-direction ($\mathbf{p}=p_{y}%
\mathbf{\hat{y}}$) with its spin pointing in the $z$-direction $\left( 
\boldsymbol{\sigma }=\mathbf{\hat{z}}\sigma _{z}\right) $ in an ensemble of randomly distributed identical point with density $n$, the expectation values $%
\langle \sigma _{z}\rangle =1$ and the average force is
\begin{equation}
\langle \mathbf{F}_{\mathrm{so}}^{yz}\rangle =\frac{n}{4\pi \epsilon _{0}}\frac{4eQ\hbar \pi^2 }{3m^{2}c^{2}} p_{y}\mathbf{\hat{x}}.
\label{eq:meanForce}
\end{equation}%
For our definition of a positive spin Hall angle we now choose the charge to be negative ($Q<0$, repulsive). We then arrive at the following right-hand rule: The electron
with its spin pointing in the $z$-direction (thumb) and moving in the $y$-direction (forefinger) drifts to the negative $x$-direction (middle finger) [Fig.~\ref{fig:defplusminus}(a)].
\red{A comparison with Eq.~\eqref{eq:Js} and using 
\begin{equation}
	\pmb{J}_{\mathrm{s},z}=-\frac{C}{e^2\rho}\nabla\mu_\mathrm{s}=\frac{C}{e^2\rho}\langle \mathbf{F}_{\mathrm{so}}^{yz}\rangle
\end{equation}
and 
\begin{equation}
	\pmb{J}_\mathrm{c}=\frac{p_y}{m}n_\mathrm{e}C\mathbf{\hat{y}}
\end{equation}
where $\mu_\mathrm{s}$ is the spin chemical potential, $\rho$ is the resistivity, $C$ is the cross sectional area the currents are flowing through, and $n_\mathrm{e}$ is the carrier density, 
%we obtain
%\begin{equation}
	%\pmb{J}_{\mathrm{s},z}=\frac{n}{4\pi \epsilon _{0}}\frac{4eQ\hbar \pi^2 }{3m^{2}c^{2}} \frac{m}{e^2\rho n_\mathrm{e}}J_\mathrm{c}\mathbf{\hat{x}}
%\label{eq:JsJc}
%\end{equation}
%where $J_\mathrm{c}=|\pmb{J}_\mathrm{c}|$. A comparison wit Eq.~\eqref{eq:Js} 
leads to a spin Hall angle
\begin{equation}
	\theta_\mathrm{SH}=-\frac{n}{4\pi \epsilon _{0}}\frac{4eQ\hbar \pi^2 }{3m^{2}c^{2}} \frac{m}{e^2\rho n_\mathrm{e}}.
\label{eq:SHA}
\end{equation}
Inserting numbers for fundamental constants $\theta_\mathrm{SH}\cong\pm\SI{3e-10}{\ohm m}\times n/(n_\mathrm{e}\rho)$ for $Q=\mp e$.}\\
%temporary location to have this fig on page 3
\begin{figure*}[tbp]
\includegraphics[width=\textwidth]{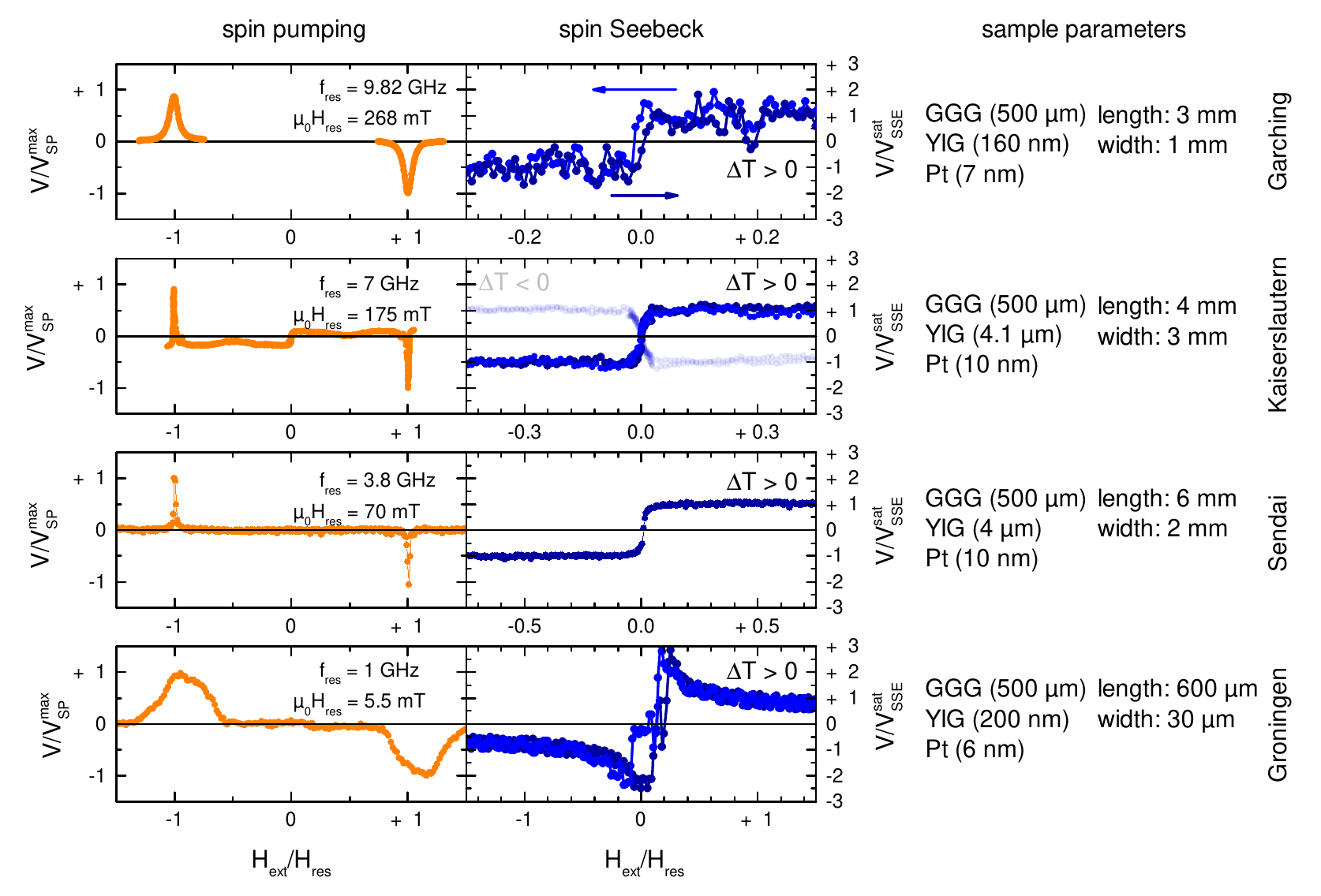}
\caption{Measured voltage signals for the FMR spin pumping (left column) and
spin Seebeck (middle column) experiments obtained by the contributing
groups. The voltage signals have been normalized to a maximum modulus of unity while the applied magnetic fields are in units of the FMR resonance field $H_\mathrm{res}$ given in the insets. The temperature difference $\Delta T=T_\mathrm{Pt}-T_\mathrm{YIG}$ is positive. The third column lists sample layer thicknesses and
dimensions. The sign of the observed voltages is consistent between the individual groups.}
\label{fig:expresults}
\end{figure*}
The time-averaged spin current injected by a steady precession~\cite%
{Tserkovnyak2002} around the equilibrium magnetization with unit vector $%
\mathbf{\hat{m}}$ is polarized along $\mathbf{\hat{m}}$. This spin pumping
process is associated with energy relaxation of the magnetization dynamics
that increases the magnetic moment in the direction of the effective
magnetic field. When the g-factor is positive, the spin pumping current
through the interface is positive as well. In the SSE, when the temperature of the
magnetization is \emph{lower} than that of the electrons in the metal, the
energy and, if $g>0$, spin current is opposite to that under FMR~\cite{Xiao2010}, leading to an opposite sign in the ISHE voltage compared to the
FMR. Under open circuit conditions \red{the inverse spin Hall effect [Eq.~\eqref{eq:ISHE}]} leads to a charge separation and an electrostatic field 
\begin{equation}
	\pmb{E}_\mathrm{s}=\frac{-e\rho}{A}\theta_\mathrm{SH}\left[\pmb{J}_{\mathrm{s},\red{\hat{\pmb{m}}}}\times\hat{\pmb{m}}\right],
\label{eq:SHE}
\end{equation}
where $A$ is the area of the ferromagnet$|$metal interface and $\rho$ is the resistivity of the metal layer. This corresponds to an electromotive force $\mathcal{E}=-\pmb{E}_\mathrm{s}\cdot\pmb{l}$, where $\pmb{l}$ is the length vector from the \textit{Lo} to the \textit{Hi} contact in Fig.~\ref{fig:defplusminus}(c) and (d).\\
The sign of an applied magnetic field is related to the current direction
according to Ampere's right hand law as depicted in Fig.~\ref{fig:defplusminus}%
(b). In practice, it is convenient to use a compass needle for comparison
with the Earth's magnetic field. Fig.~\ref%
{fig:defplusminus}(b) defines the positive field direction 
from the antarctic to the arctic, i.e. along the geomagnetic
field. Typical experimental setups for spin pumping [(c)] and spin Seebeck
experiments [(d)] on yttrium iron garnet$|$platinum thin film (YIG$|$Pt)
bilayers are also sketched in Fig.~\ref{fig:defplusminus}. In the former, a
ferromagnet (F)$|$normal metal (N) stack is exposed to microwave radiation
with frequency $f$ (typically in the GHz regime), while in spin
Seebeck experiments the bilayer is exposed to a thermal gradient. Sample
parameters used by the different groups are listed in the third column of
Fig.~\ref{fig:expresults}. For details on the sample fabrication we refer to
Refs.~\onlinecite{Geprags2012,Althammer2013} (WMI), Ref.~%
\onlinecite{Castel2012} (RUG), Refs.~%
\onlinecite{Jungfleisch2013,
Jungfleisch2013a} (UniKL) and Ref.~\onlinecite{Qiu2013} (IMR). 

Fig.~\ref{fig:expresults} summarizes the results of the participating
groups. Note that in each group both the spin pumping and spin Seebeck experiments were performed on the same sample, without changing the setup.\\

At the WMI, FMR experiments (first row) were carried out in a
microwave cavity with fixed frequency $f_{\mathrm{res}}=\SI{9.82}{\giga Hz}$ as a
function of an applied magnetic field $H_{\mathrm{ext}}$ leading to
resonance at $\mu _{0}H_{\mathrm{ext}}\cong \SI{270}{\milli T}$. A source meter was used
to drive a large ($I_{\mathrm{h}}=\pm \SI{20}{\milli A}$) dc charge current through the platinum film
($R_{\mathrm{Pt}}=\SI{197}{\ohm}$) in order to generate a temperature
gradient (hot Pt, cold YIG)~\cite{Schreier2013a}. By summing voltages recorded for opposite $I_{%
\mathrm{h}}$ directions, the magnetoresistive contributions cancel out, such
that only the spin Seebeck signal remains. The ISHE voltages for both FMR and the spin Seebeck
experiments were measured by the same, identically connected nanovoltmeter with
microwave and heating current separately turned on.

Results from the UniKL are shown in the second row in Fig.~\ref{fig:expresults}. A
microwave with $f_{\mathrm{res}}=\SI{7}{\giga Hz}$ fed into a Cu stripline on top of
the Pt film excites the FMR at an external magnetic field of $\mu _{0}H_{\mathrm{%
ext}}\cong\SI{175}{\milli T}$. The microwave current amplitude was modulated at a
frequency of $f_{\mathrm{mod}}=\SI{500}{Hz}$ to allow for lock-in detection of the
induced voltages~\cite{Jungfleisch2013a} that are measured by a nanovoltmeter. The spin pumping data show a small
offset between positive and negative magnetic fields, which stems from Joule heating in the Pt layer by the microwaves. Peltier elements on
the top and bottom (separated by an AlN layer) generated a thermal gradient
for the spin Seebeck experiments that were reversed for cross checks, as
shown in the right graph.

The third row in Fig.~\ref{fig:expresults} shows the results obtained at
the IMR. Here, the sample is placed on a coplanar \mbox{waveguide} such that at $f_{%
\mathrm{res}}=\SI{3.8}{\giga Hz}$ FMR condition is fulfilled for $%
\mu _{0}H_{\mathrm{ext}}\cong\SI{70}{\milli T}$. The thermal gradient for the spin Seebeck measurements was generated by an electrically isolated separate heater on top of the Pt.

The fourth row in Fig.~\ref{fig:expresults} shows the RUG results. A coplanar
waveguide on top of the YIG was used to excite the FMR at a magnetic field of 
$\mu _{0}H_{\mathrm{ext}}\cong \SI{6}{\milli T}$. The spin Seebeck effect was detected
using an ac-variant of the current heating scheme. The spin Seebeck
voltage can thereby be detected as described above in the second harmonic of the ac voltage signal.

In spite of differences in samples and measurement techniques, all
experiments agree on the sign for spin pumping and spin Seebeck effect. We
all measure \textit{negative} spin pumping and \textit{positive} spin
Seebeck voltages for positive applied magnetic fields that all change sign
when the magnetic field is reversed, consistent with the theoretical
expectations~\cite{Dyakonov1971,Hirsch1999,Tserkovnyak2002, Xiao2010}.

We can now address the \emph{absolute} sign of the spin Hall angle. 
The results in Fig.~\ref{fig:expresults} were obtained with measurement
configurations equivalent to the one depicted in Figs.~\ref{fig:defplusminus}%
(c) and (d).
With external magnetic field $\mathbf{H}_{\mathrm{ext}}$ pointing in the 
$\mathbf{\hat{z}}$ direction $\mathbf{\hat{m}}=\mathbf{\hat{z}}$ and
$\hat{\pmb{J}}_{\mathrm{s}}=\hat{\pmb{y}}$ for FMR spin pumping. According to Eq.~\eqref{eq:SHE}, when $%
\theta _{\mathrm{SH}}>0$ $\hat{\pmb{E}}_{\mathrm{s}}=-\mathbf{\hat{x%
}}$, which leads to a negative (positive) charge accumulation at the $-x$ $\left( +x\right) $
edge of the Pt film and a negative spin
pumping voltage is expected as well as observed. In the spin Seebeck
experiments with Pt hotter than YIG, the spin current flows in the
opposite direction ($\hat{\pmb{J}}_{\mathrm{s}}=-\mathbf{\hat{y}}$%
), and the voltage is inverted. Therefore, the spin Hall angle of Pt is
positive if defined as above.
The nature of the spin Hall effect in Pt is likely to be governed by its
electronic band structure~\cite{Guo2008}, but it should be a helpful to know that the sign is identical to that caused by negatively charged impurities.\\

In summary, we present spin pumping and spin Seebeck experiments for various
samples and experimental conditions leading to gratifying agreement of the
results obtained by different groups. By carefully accounting for the signs
of all experimental parameters and definitions we were able to determine
both the relative and the absolute signs of both effects, linking the positive spin Hall angle of Pt to a simple physical model of negative scattering centers. The relative signs of spin pumping and spin Seebeck
effect are consistent with theoretical predictions~\cite{Tserkovnyak2002a,
Xiao2010, Adachi2011, Hoffman2013}. The techniques and samples used in this letter are representative for a large number of spin pumping and spin Seebeck experiments and should serve as a reference for other materials or sample geometries.\\ 

We thank M. Wagner, M. Althammer and M. Opel for sample preparation and gratefully acknowledge financial support by the DFG via SPP 1538 ``Spin
Caloric Transport'' (projects GO 944/4-1, SE 1771/4-1, BA 2954/1-1) and CH 1037/1-1, NanoLab NL and the Foundation for
Fundamental Research on Matter (FOM), \red{JSPS} Grants-in-Aid for
Scientific Research, EU-RTN Spinicur, EU-FET InSpin 612759, PRESTO-JST ``Phase Interfaces for Highly Efficient Energy Utilization'', and CREST-JST ``Creation of Nanosystems with Novel Functions through Process Integration''.

\end{document}